\documentclass[11pt]{article}

\usepackage{amsmath,amssymb}
\usepackage{graphicx,authblk}
\newcommand{\amp}{&\!\!}

\newcommand{\beq}{\begin{equation}}
\newcommand{\eeq}{\end{equation}}
\newcommand{\bea}{\begin{eqnarray}}
\newcommand{\eea}{\end{eqnarray}}

\begin{document}

\title{\LARGE{\bf Entanglement Entropy in Scalar Field Theory }}

\author{
~\\
{\Large Mark P. Hertzberg} \\
~\\
{\em Center for Theoretical Physics, Department of Physics},\\
{\em Massachusetts Institute of Technology, Cambridge, MA 02139, USA}\\
~\\
{\em Department of Physics},\\ 
{\em Stanford University, Stanford, CA 94305, USA}}

\date{}

\maketitle 

\begin{abstract}
Understanding the dependence of entanglement entropy on the renormalized mass in quantum field theories can provide insight into phenomena such as quantum phase transitions, since the mass varies in a singular way near the transition.
Here we perturbatively calculate the entanglement entropy in interacting scalar field theory, focussing on the dependence on the field's mass. We study $\lambda\phi^4$ and $g\phi^3$ theories in their ground state. By tracing over a half space, using the replica trick and position space Green's functions on the cone, we show that space-time volume divergences cancel and renormalization can be consistently performed in this conical geometry.
We establish finite contributions to the entanglement entropy up to two-loop order, involving a finite area law. 
The resulting entropy is simple and intuitive: the free theory result in $d=3$ (that we included in an earlier publication) $\Delta S\sim A\,m^{2}\ln(m^2)$ is altered, to leading order, by replacing the bare mass $m$ by the renormalized mass $m_r$ evaluated at 
the renormalization scale of zero momentum.
\let\thefootnote\relax\footnotetext{Electronic address: {\tt mphertz@mit.edu}}
\end{abstract}

\vspace*{-15.9cm} {\hfill MIT-CTP 4395\,\,\,\,}



\newpage


\section{Introduction}

Entanglement entropy is an important property of quantum mechanical states. 
By tracing out a region of a system, the resulting von Neumann or entanglement entropy $S=-\mbox{Tr}[\rho\ln\rho]$ is a measure of one's inability to describe the full system if one only has access to a subsystem. There are other ways to characterize entanglement, but the entropy is particularly fruitful. This fundamental object appears in various contexts, including quantum field theory, condensed matter physics, black hole thermodynamics, holography, and other regimes in which quantum mechanics plays a central role. For instance, at so-called quantum phase transitions, the entanglement entropy can act as a diagnostic for the nature of the phase transition. Entanglement entropy has been the subject of various recent investigations (including 
Refs.~\cite{Sachdev2009,Casini:2009sr,Ryu:2006ef,Nishioka:2009un,Fursaev:2006ng,Eisert:2008ur,Solodukhin:2011gn,Song,Ding,Doyon}).

A central feature of the entanglement entropy of fields in spatial dimensions $d>1$ is that the entropy is strictly cutoff dependent (unlike the $d=1$ case which involves only a logarithm \cite{Holzhey:1994we,Callan:1994py}).
In $d=3$ it was shown in Ref.~\cite{Srednicki:1993im} (and earlier work in \cite{Bombelli}) that for a scalar field in its ground state the entanglement entropy between the interior and exterior of a sphere is 
\beq
S\sim {A\over\epsilon^2}+\mbox{subleading}
\label{SArea}\eeq
where $A$ is the surface area of the sphere and $\epsilon$ is some microscopic cutoff on the field theory. (This area law can be altered to include logarithmic corrections for more complex field theories, such as those involving fermions \cite{Swingle:2009}.) It is important to note that the entanglement entropy is necessarily cutoff dependent and sensitive to the detailed microphysics. For instance, in a lattice field theory it would depend on the lattice spacing and arrangement.
Instead if one wishes to use the entanglement entropy to characterize UV cutoff independent properties of a system, such as the IR behavior at or near a quantum phase transition, one needs to establish such sub-leading corrections to the entropy. In fact such contributions do exist,
as has been examined in the literature (e.g., see \cite{Casini:2004bw,Casini:2006hu,Kitaev:2005dm,Hung:2011ta}).

Recently in Ref.~\cite{Hertzberg:2010uv} we studied free scalar field theory in $D=d$+1 space-time dimensions.
The entanglement entropy was computed for a collection of geometries, including a waveguide with various boundary conditions.
In addition to the usual cutoff dependent area piece $S_{div}\sim A_{\perp}/\epsilon^{d-1}$, where $A_\perp$ is the $d-1$ dimensional area of the transverse space and $\epsilon$ is a microscopic cutoff, there is also a finite and cutoff independent area law
\bea
&&\Delta S=
\bigg{\{}
\begin{array}{c}
\!\gamma_d \, A_{\perp}\,m^{d-1}\ln(m^2),\,\,\,\,\mbox{for}\,\,\,d\,\,\,\mbox{odd}\\
\gamma_d \, A_{\perp}\,m^{d-1},\,\,\,\,\,\,\,\,\,\,\,\,\,\,\,\,\,\,\,\,\,\,\,\mbox{for}\,\,\,d\,\,\,\mbox{even}
\label{FreeOddEven}\end{array}
\eea
where $\gamma_d\equiv (-1)^{(d+1)/2}[12\,(4\pi)^{(d-1)/2}((d-1)/2)!]^{-1}$ for an odd number of spatial dimensions $d$
and $\gamma_d\equiv(-1)^{d/2}[12\,(2\pi)^{(d-2)/2}(d-1)!!]^{-1}$ for an even number of spatial dimensions $d$.  This was computed in Ref.~\cite{Hertzberg:2010uv} using heat kernel methods, which were earlier used in Ref.~\cite{Callan:1994py}. There was also found further sub-leading contributions, including a perimeter term and a curvature, or topological, term.
These contributions are interesting in the sense that not only does this constitute a UV cutoff independent contribution that scales as the area, it also varies with the field's mass. In the vicinity of a critical point, the mass varies in a singular fashion, and therefore so does this contribution to the entropy (by ``singular" we mean that derivatives diverge near the transition, though the mass itself approaches zero). So in principle, such a contribution can discriminate between phases. 

It is important to note that our previous results were computed only in the very special case of free field theory. It is natural then to ask to what extent do such results persist in interacting field theories.
So here we would like to extend these results to the interacting case. In particular, we will focus on scalar field theories of the form
\beq
\mathcal{L}={1\over2}(\partial\phi)^2-{1\over2}m^2\phi^2-{g\over3!}\phi^3-{\lambda\over4!}\phi^4
\label{Lagrangian}\eeq
i.e., renormalizable theories in $d\leq 3$, and will study them perturbatively for small $g,\,\lambda$.
We are especially interested in $d\leq 3$ here, so we can track {\em all} the various divergences that emerge in the calculation. For $d>3$, the theory is non-renormalizable, and one would need to track new divergences. In the latter case, our results would still properly account for the {\em leading} divergences, but there may be sub-leading ones.

For definiteness, consider the $d=3$ interacting case. Here $\lambda$ is dimensionless and $g$ has units of mass ($\hbar=c=1$).
This implies that it would be dimensionally correct to have area law corrections to eq.~(\ref{SArea}) of the form
\beq
S_1\sim \lambda{A\over\epsilon^2},\,\,\,\,\mbox{or}\,\,\,\,S_1\sim g^2 A \ln\epsilon,\,\,\,\,\mbox{etc}
\label{fakes}\eeq
(possibly multiplied by further dimensionless logarithmic factors), where the ``1" subscript indicates that these are leading order in the couplings.
If such contributions did exist, it would compromise the importance of the previously stated results of eq.~(\ref{FreeOddEven}).
Recall that the significance of the free theory result in (\ref{FreeOddEven}) is that they are manifestly cutoff independent. This means that as the IR parameters of the field theory vary, say in the vicinity of a critical point, then the cutoff dependent piece $\sim A/\epsilon^2$ would drop out, leaving the purely mass dependent and cutoff independent contributions. On the other hand, if in the interacting theory, contributions such as that suggested on dimensional grounds in eq.~(\ref{fakes}) existed, then as one varies the IR parameters, such as $\lambda$ or $g$, the entropy would vary in an explicitly cutoff {\em dependent} way, due to the $\epsilon$ dependence. 

In this work we show, to two-loop order, that the contributions of the form (\ref{fakes}) in fact do exist when written in terms of the {\em bare} parameters, in addition to the contributions in (\ref{SArea},\,\ref{FreeOddEven}). However, the main result of this work is that we show, once renormalization has been consistently performed, the contributions in (\ref{fakes}) are absorbed into {\em flat space renormalization of the field's mass}. The primary change from the free theory result is that the entropy in (\ref{FreeOddEven}), written in terms of the bare mass $m$, is replaced by the same result in terms of the renormalized mass $m_r$ evaluated at the renormalization scale of zero momentum. So as one varies the IR parameter $m_r$, there is a calculable piece that changes, while the UV cutoff dependent pieces do not.
This leaves a cutoff independent area law contribution to the entropy, even in the interacting case.

The organization of this paper is as follows: 
In Section \ref{DensMat} we introduce the  density matrix and partition function on the cone, as is demanded by the so-called replica trick. 
In Section \ref{Green} we study the Green's function on the cone. In Section \ref{Free} we recapitulate the free theory result.
In Section \ref{Quartic} we study $\lambda \phi^4$ theory to leading non-trivial order in $\lambda$. 
In Section \ref{MassRen} we connect the results to mass renormalization in flat space. 
In Section \ref{Cubic} we extend these results to $g\phi^3$ theory. 
Finally, in Section \ref{Discussion} we discuss our results.

\section{Density Matrix and Partition Function}\label{DensMat}

\begin{figure}[t]
\center{\includegraphics[width=6cm]{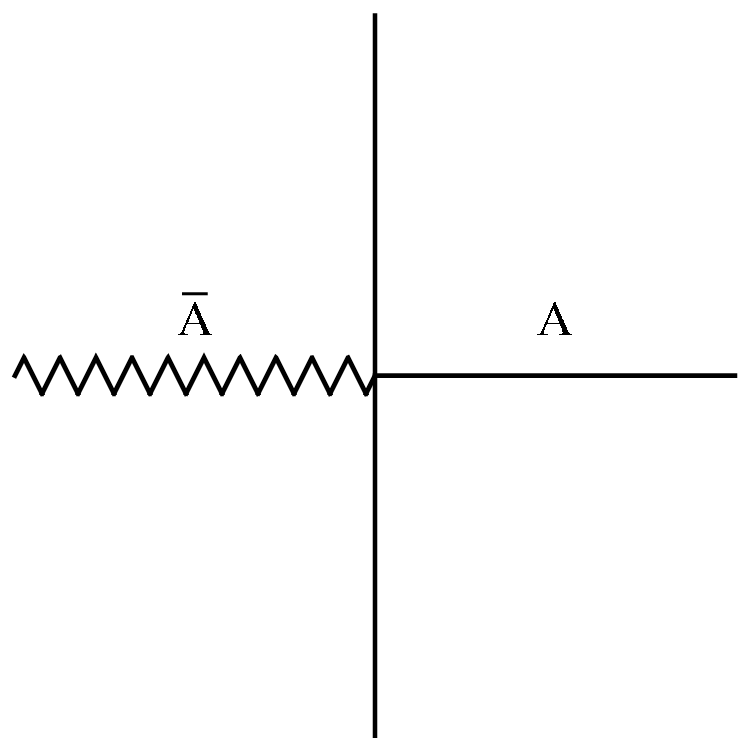}\,\,\,\,\,\,\,\,\,\,\,\,\,\,\,\,\,\,\,\,\,\,\,\,\,\,\,\,\,\,\,
\includegraphics[width=6cm]{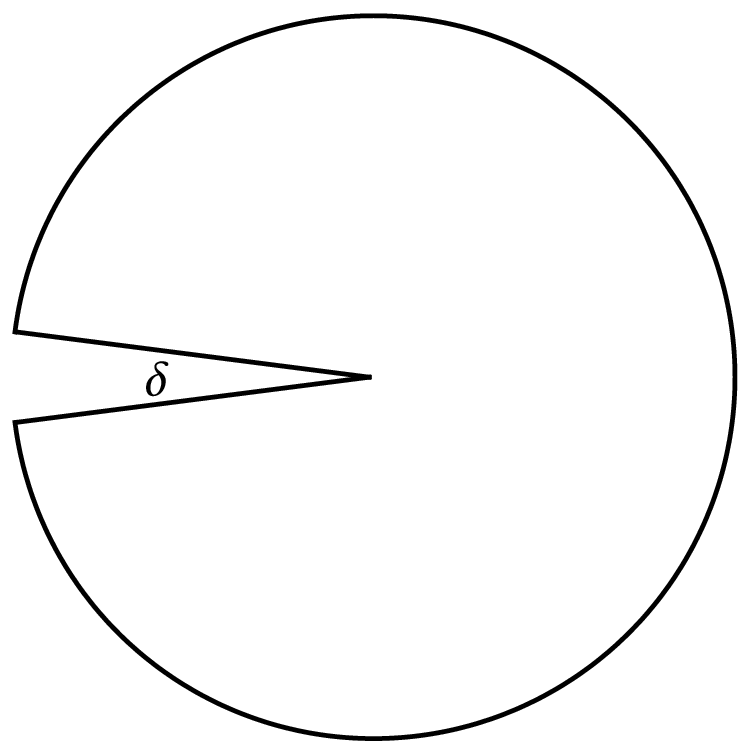}}
\caption{Tracing over the left half space $\bar{\mbox{A}}$ requires a cut along the negative real axis (left) and then a computation on the $n$th Riemann sheet in order to compute $\mbox{Tr}[\rho^n]$ (the vertical axis is Euclideanized time). For $n$ near 1 we can perform a computation on the cone (right) with deficit angle $\delta=2\pi(1-n)$, (the cone's radius is sent to infinity at the end of computation.) }
\label{Replica}\end{figure}

The fundamental starting point for the study of entanglement entropy is the full density matrix of a system $\hat{\rho}$. When we only have access to a subsystem A,  we focus on the reduced density matrix $\rho=\mbox{Tr}_{\tiny{\bar{\mbox{A}}}}\!\left[\hat{\rho}\right]$ after tracing out the complimentary region $\bar{\mbox{A}}$. In this paper we shall focus on the basic problem where A and $\bar{\mbox{A}}$ represent half-spaces. Their dividing boundary is a flat space of  dimension 
$d_\perp=d-1$ (counting spatial dimensions). Although other geometries are of interest, this is normally the basic starting point since any smooth geometry is locally flat.
Various physical quantities can be computed from the final trace over region A of some power of $\rho$. For example, the Renyi and Tsallis entropies are defined as
\beq
R(n)={\ln\mbox{Tr}\left[\rho^n\right]\over 1-n},\,\,\,\,\,\,T(n)={\mbox{Tr}\left[\rho^n\right]-1\over 1-n}
\eeq
(where  ``Tr" means ``Tr$_{\tiny{\mbox{A}}}"$ here).
These quantities are normally computed for integer values $n=2,3,\ldots$. The trace of the $n$th power can be obtained by introducing a Euclideanized time variable, then forming a cut over the traced out region $\bar{\mbox{A}}$ and performing a computation on the $n$th Riemann sheet; see Fig.~\ref{Replica} (left).
For the $n=1$ case we need to lift $n$ to be non-integer and take the $n\to1$ limit. Using L'Hopital's rule this may be written as
\beq
S=-{\partial\over\partial n}\ln\mbox{Tr}\left[\rho^n\right]\Bigg{|}_{n\to1}=-\mbox{Tr}[\rho\ln\rho]
\eeq
which is the so-called replica trick.
$S$ is the entanglement entropy and will be our focus, though other entropies (such as Renyi and Tsallis) will be readily attainable with our methods.

By the replica trick, the $n$th power of the reduced density matrix can be obtained from defining the field theory on a cone with deficit angle $\delta=2\pi(1-n)$; see Fig.~\ref{Replica} (right). We will denote all quantities on the cone with subscript $n$, including the partition function $\mathcal{Z}_n$  ($n\to1$ is flat space). For the field in its ground state, the $n$th power of the reduced density matrix is given by
\beq
\ln\mbox{Tr}\left[\rho^n\right]=\ln\mathcal{Z}_n-n\ln\mathcal{Z}_1
\eeq
and the entanglement entropy can be expressed as
\beq
S=-{\partial\over\partial n}\Big{[}\ln\mathcal{Z}_n-n\ln\mathcal{Z}_1\Big{]}\Bigg{|}_{n\to1}
\eeq
Hence our goal will be to compute the partition function $\mathcal{Z}_n$, which we shall do so perturbatively in powers of the coupling
\beq
\ln\mathcal{Z}_n=\ln\mathcal{Z}_{n,0}+\ln\mathcal{Z}_{n,1}+\ldots
\eeq

At zero temperature the partition function on the cone is given by
\beq
\mathcal{Z}_n=\int\mathcal{D}\phi \, e^{-S_E[\phi]}
\eeq
where we have Euclideanized the action.
For example, the free theory (written in terms of the bare mass $m$ and bare field $\phi$)
has the formal solution
\bea
\ln\mathcal{Z}_{n,0}\amp=\amp-{1\over2}\ln\mbox{Det}\left[-\Delta+m^2\right]
\eea
The derivative of this quantity, with respect to $m^2$, can be readily expressed in terms of an integral over the Green's function at coincidence
\beq
{\partial\over\partial m^2}\ln\mathcal{Z}_{n,0}=-{1\over2}\int_n d^Dx\,G_n({\bf x},{\bf x})
\label{Zderiv}\eeq
Indeed all needed quantities can be expressed as integrals over powers of the Green's function on the cone $G_n$; a quantity that we examine in the next section.

\section{Green's Function}\label{Green}

Previously, in the case of the free theory, the entire analysis could be performed in terms of the heat kernel (or the density of states) on the cone, 
but the interacting case will require going to the Green's function as it carries more information.

So to make progress perturbatively, we need the Green's function on the cone $G_n$.
The Green's function satisfies the differential equation
\beq
(-\Delta+m^2)G_n({\bf x},{\bf x}')=\delta^D({\bf x}-{\bf x}')
\eeq
It is important to note that since the tip of the cone breaks translational invariance, then $G_n({\bf x},{\bf x}')$ is a function of both ${\bf x}$ and ${\bf x}'$ separately, even though it is only be a function of the distance $|{\bf x}-{\bf x}'|$ in flat space.
Let us write our co-ordinates as ${\bf x}=\{r,\theta,{\bf x}_\perp\}$, where $\{r,\theta\}$ are the polar co-ordinates on the cone and ${\bf x}_\perp$ are co-ordinates on the $d_\perp=d-1$ dimensional transverse space. The solution for the Green's function can be expressed as the following integral over transverse momentum and summation (e.g., see Ref.~\cite{Calabrese:2004eu})
\beq
G_n({\bf x},{\bf x}')={1\over2\pi n}\int\!{d^{d_\perp}\!p_\perp\over(2\pi)^{d_\perp}}\sum_{k=0}^\infty d_k\int_0^\infty dq\,q{J_{k/n}(q\, r)J_{k/n}(q\, r')\over q^2+m^2+p_\perp^2}\cos(k(\theta-\theta')/n)
\,e^{i\,{\bf p}_\perp\cdot({\bf x}_\perp-{\bf x}_\perp')}
\eeq
where $J$ is the Bessel function of the first kind, and the coefficients are $d_{k=0}=1,\,d_{k\geq1}=2$.

Although the full expression for $G_n$ for arbitrary ${\bf x}$ and ${\bf x}'$ is required to analyze the full theory, it is somewhat unwieldy.
We shall return to this in Section \ref{Cubic}, where it will be essential. However, for the purposes of the free theory and the $\lambda\phi^4$ theory at leading order in $\lambda$, it will suffice to study the coincidence limit ${\bf x}'\to {\bf x}$. In this limit there are simplifications to the form of the Green's function.
Indeed by setting ${\bf x}'={\bf x}$ we can carry out the $q$ integral, with the result
\beq
G_n({\bf x},{\bf x})={1\over 2\pi n}\int\!{d^{d_\perp}\!p_\perp\over(2\pi)^{d_\perp}}\sum_{k=0}^\infty d_k \,
I_{k/n}\Big(\sqrt{m^2+p_\perp^2}\, r\Big)K_{k/n}\Big(\sqrt{m^2+p_\perp^2}\, r\Big)
\eeq
where $I$ and $K$ are the modified Bessel function of the first and second kind, respectively.

Following Ref.~\cite{Calabrese:2004eu} we now employ the Euler-Maclaurin formula
\beq
\sum_{k=0}^\infty d_k F(k)=2\int_0^\infty \!dk\,F(k)-{1\over6}F'(0)
-2\sum_{j>1}^\infty{B_{2j}\over(2j)!}F^{(2j-1)}(0)
\eeq
(where $B_{2j}$ are the Bernoulli numbers)
to formally re-express the sum over $k$ in terms of an integral over $k$ plus a sum over a dummy index $j$. This leads to
\bea
G_n({\bf x},{\bf x})={1\over 2\pi n}\int\!{d^{d_\perp}\!p_\perp\over(2\pi)^{d_\perp}}\Bigg{[}2\!\int_{0}^\infty dk \,
I_{k/n}\Big(\sqrt{m^2+p_\perp^2}\, r\Big)K_{k/n}\Big(\sqrt{m^2+p_\perp^2}\, r\Big)\nonumber\\
+{1\over 6n}K_0^2\Big(\sqrt{m^2+p_\perp^2}\,r\Big)\Bigg{]}+(j>1)
\eea
Here the $j>1$ contributions to the summation do not lead to divergences at the order we are working, and do not concern us here.
In the first term we can rescale the dummy variable of integration to $\bar{k}=k/n$ in order to extract the $n$-dependence. This allows us to re-express the integral in terms of $G_1$, namely
\beq
G_n({\bf x},{\bf x})=G_1(0)+f_n(r)
\label{GreenCoinc}
\eeq
where we have introduced the function
\beq
f_n(r)\equiv{1\over2\pi n}{1-n^2\over 6n}\int\!{d^{d_\perp}\!p_\perp\over(2\pi)^{d_\perp}}K_0^2\Big(\sqrt{m^2+p_\perp^2}\,r\Big)+(j>1)
\label{GreenCoinc2}
\eeq
The first term on the right hand side in eq.~(\ref{GreenCoinc}) is the usual flat space Green's function, which is formally divergent at coincident points. Since the flat space Green's function carries translational invariance, it is only a function of $|{\bf x}-{\bf x}'|$, giving 
\beq
G_1({\bf x},{\bf x})=G_1(|{\bf x}-{\bf x}|)=G_1(0).
\eeq
The second term in eq.~(\ref{GreenCoinc}) is finite and depends on radius (distance from the tip of cone). Since $K_0$ decays exponentially fast at large $r$, the Green's function on the cone $G_n$ rapidly approaches the Green's function on flat space $G_1$ at large $r$, as expected.

\section{Free Theory Result}\label{Free}

For the free theory (or zeroth order in the couplings in the interacting theory)
we can use the expression in eq.~(\ref{Zderiv}) to determine the derivative of the density matrix as follows
\bea
{\partial\over\partial m^2}\ln\mbox{Tr}[\rho^n_{\,0}]\amp=\amp{\partial\over\partial m^2}\ln\mathcal{Z}_{n,0}
-n{\partial\over\partial m^2}\ln\mathcal{Z}_{1,0}\\
\amp=\amp-{1\over2}\left[\int_{n} d^Dx\,G_n({\bf x},{\bf x})-n\int d^D x\,G_1(0)\right]
\eea
where the first integral is over the cone (as implied by the $n$ subscript on the integral sign) and the second integral is over flat space; so the angular integral gives a factor of $2\pi n$ in the first integral and a factor of $2\pi$ in the second integral.
Substituting in (\ref{GreenCoinc}), we see that the $G_1$ term cancels. This is very important because such term would otherwise scale with the space-time volume multiplied by a UV divergent factor. Such space-time volume divergences are incompatible with the area law scaling and should always cancel out; a point we will return to later.

After the cancellation, we are left with
\beq
{\partial\over\partial m^2}\ln\mbox{Tr}[\rho^n_{\,0}]=-{1-n^2\over 12n}\int d^{d_\perp}\!x_\perp\int_0^\infty dr\,r
\int\!{d^{d_\perp}\!p_\perp\over(2\pi)^{d_\perp}}K_0^2\Big(\sqrt{m^2+p_\perp^2}\,r\Big)
\eeq
The integral over $r$ can be easily performed using $\int_0^\infty dy\,y\, K_0^2(y)=1/2$, and the integral over the transverse space is $\int d^{d_\perp}\!x_\perp=A_\perp$, giving
\beq
{\partial\over\partial m^2}\ln\mbox{Tr}[\rho^n_{\,0}]=-{1-n^2\over24n}A_\perp\!\int\!{d^{d_\perp}\!p_\perp\over(2\pi)^{d_\perp}}{1\over m^2+p_\perp^2}
\eeq
Then by integrating this with respect to $m^2$, and performing the $n$ derivative, we are able to obtain the free contribution to the entropy.
For the entanglement entropy we have
\beq
S_0=-{1\over12}A_\perp\!\int\!{d^{d_\perp}\!p_\perp\over(2\pi)^{d_\perp}}\ln(m^2+p_\perp^2)+\mbox{const}
\label{FreeEntropy}\eeq
where ``const" represents a (UV divergent) quantity, independent of mass.
By performing the integral over the transverse momenta $p_\perp$ in different dimensions, and extracting only the finite pieces, we recover the free theory results that we reported in Ref.~\cite{Hertzberg:2010uv} and recapitulated in the introduction (\ref{FreeOddEven}).

\section{Quartic Interaction}\label{Quartic}

With the free theory result in hand, we can now perturbatively compute corrections from interactions. To do so we return to the full partition function.
For the quartic $\lambda\phi^4$ theory, the Euclidean partition function can be expanded to order $\lambda$ as
\bea
\ln\mathcal{Z}_n\amp=\amp\ln\int\mathcal{D}\phi\,e^{-S_E[\phi]}\\
\amp=\amp \ln\mathcal{Z}_{n,0}-{\lambda\over 4!}\int_n d^Dx\,\left\langle\phi({\bf x})^4\right\rangle_0+\ldots\\
\amp=\amp \ln\mathcal{Z}_{n,0}-{3\lambda\over4!}\int_n d^Dx\,G_n({\bf x},{\bf x})^2+\ldots
\eea
where $\mathcal{Z}_{n,0}$ is the partition function of the quadratic theory, and $G_n$ is the Green's function evaluated at coincident points; both quantities are defined on the cone, the subscript $n$ on the integral indicates that it is to be performed over the cone, and the factor of 3 comes from an application of Wick's theorem.

Here we need the square of the Green's function, which we write as
\beq
G_n({\bf x},{\bf x})^2=G_1(0)^2+2G_1(0)f_n(r)+f_n(r)^2.
\eeq
So the $\mathcal{O}(\lambda)$ contribution to the density matrix is
\bea
\ln\mbox{Tr}[\rho^n_{\,1}]\amp=\amp\ln\mathcal{Z}_{n,1}-n\ln\mathcal{Z}_{1,1}\\
\amp=\amp-{3\lambda\over4!}\left[\int_{n} d^Dx\,G_n({\bf x},{\bf x})^2-n\int d^D x\,G_1(0)^2\right]\\
\amp=\amp-{3\lambda\over 4!}A_\perp 2\pi n\int_0^\infty dr\,r\left[2G_1(0)f_n(r)+f_n(r)^2\right]\\
\amp=\amp-{3\lambda\over4!}A_\perp \left[{1-n^2\over 6n}G_1(0) \int{d^{d_\perp}\!p_\perp\over(2\pi)^{d_\perp}}{1\over m^2+p_\perp^2} +2\pi n\int_0^\infty dr\,r\, f_n(r)^2\right]
\eea
(again dropping the $j>1$ sub-leading terms). We note that the final term here vanishes when we differentiate with respect to $n$ and then set $n=1$, so the only term that contributes
is the first term. This gives the result
\beq
S_1=-{\lambda\over 4!}A_\perp G_1(0)\int\!{d^{d_\perp}\!p_\perp\over(2\pi)^{d_\perp}}{1\over m^2+p_\perp^2} 
\eeq
Consider the $d=3$ example ($d_\perp=2$). In this case, the integral is dimensionless and does not lead to any power law divergences. 
So we have $S_1\sim \lambda\, A \, G_1(0)$ (up to logarithmic corrections). In $3+1$-dimensions, the Green's function has a quadratic divergence at coincidence, which we can regulate in position space with a microscopic cutoff $\epsilon$, giving $G_1(0)\sim 1/\epsilon^2$. Hence, $S_1\sim \lambda\,A/\epsilon^2$, as anticipated in the introduction.

By combining this with the free theory result in (\ref{FreeEntropy}), we have the following total result for the entropy at $\mathcal{O}(\lambda)$
\bea
S\amp=\amp S_0+S_1+\ldots\\
\amp=\amp-{1\over 12}A_\perp \int\!{d^{d_\perp}\!p_\perp\over(2\pi)^{d_\perp}}
\left[\ln(m^2+p_\perp^2)+{\lambda\over 2}{G_1(0)\over m^2+p_\perp^2} \right]+\mbox{const}+\ldots
\eea
For a chosen number of transverse dimensions $d_\perp$ the integral over $p_\perp$ can be readily performed. However, an understanding of the divergences in this result, coming from the $G_1(0)$ factor and the high $p_\perp$ regime, is quite unclear when left in this form.

\section{Mass Renormalization} \label{MassRen}

In order to understand the structure of the previous result, in particular the organization of divergences, it is important to re-express the result in terms of {\em renormalized} couplings, rather than the {\em bare} couplings that we have studied so far. 

\begin{figure}[t]
\center{\includegraphics[width=3cm]{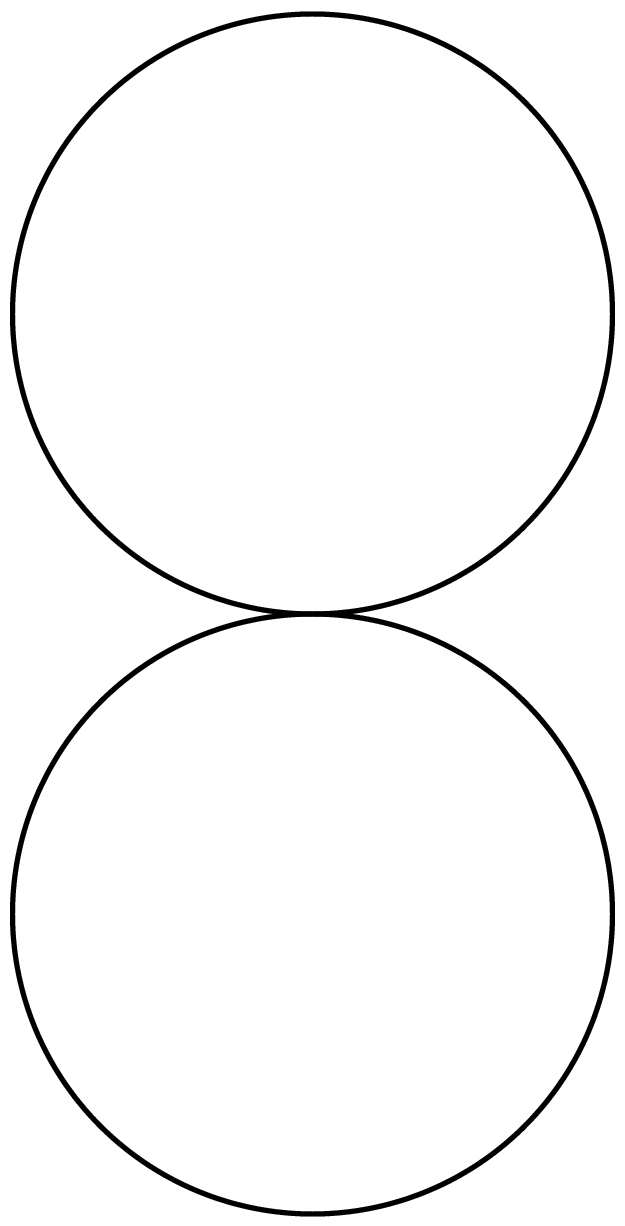}\,\,\,\,\,\,\,\,\,\,\,\,\,\,\,\,\,\,\,\,\,\,\,\,\,\,\,\,\,\,\,\,\,\,\,\,\,\,\,
\includegraphics[width=6cm]{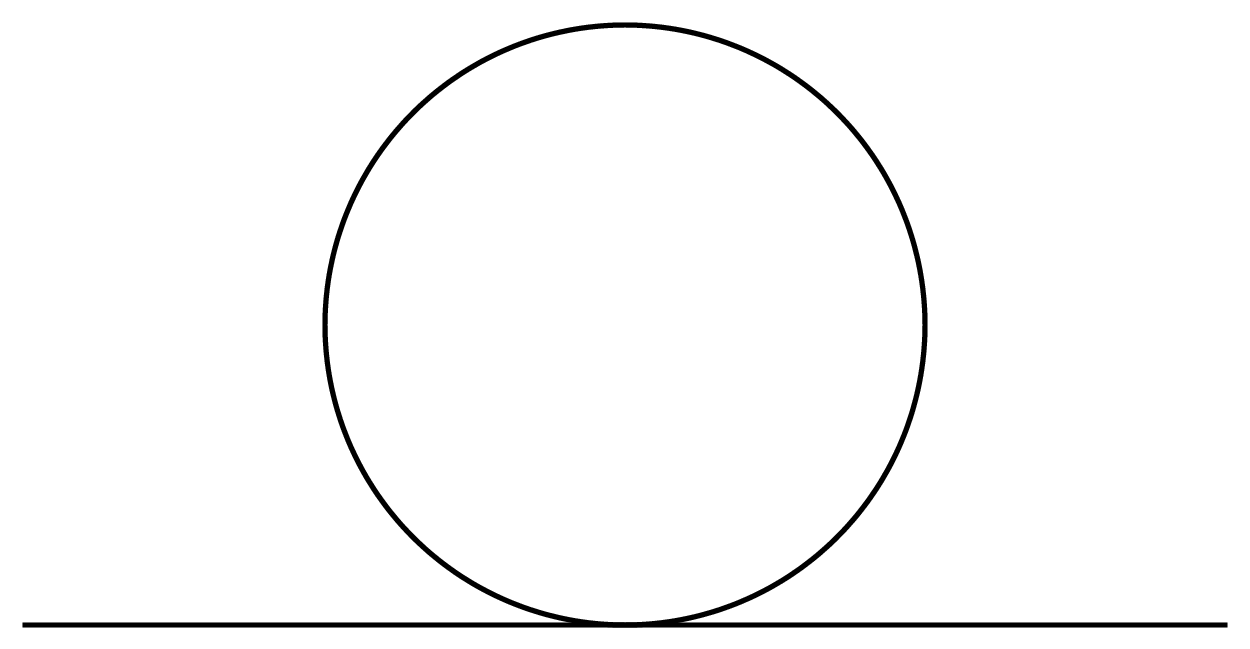}}
\caption{Feynman diagrams for $\phi^4$ theory at order $\lambda$: two-loop vacuum contribution to entropy (left) and one-loop contribution to mass renormalization (right).}
\label{Phi4}\end{figure}

Since the UV divergences arise from arbitrarily short distance physics, and since (away from the tip of the cone) 
the field theory on the cone recovers the field theory in flat space, then we should recall {\em flat space renormalization}.
Recall that the renormalized mass $m_{r}^2$ at one-loop in flat space $\lambda\phi^4$ theory is given in terms of the bare mass $m^2$ by
\bea
m_{r}^2\amp=\amp m^2+{\lambda\over2}\int\!{d^D p\over(2\pi)^D}\,\hat{G}_1(p)\\
\amp=\amp m^2+{\lambda\over2}G_1(0)
\eea
Let us conjecture that the full result for the entropy, or at least to the order we are working, takes the result from the free theory $S_0$ in (\ref{FreeEntropy}) with the bare mass replaced by the renormalized mass $m_r$, i.e.,
\beq
S=-{1\over12}A_\perp\!\int\!{d^{d_\perp}\!p_\perp\over(2\pi)^{d_\perp}}\ln(m_{r}^2+p_\perp^2)+\mbox{const}
\label{FullEntropy}\eeq
By expanding this to $\mathcal{O}(\lambda)$ we indeed recover exactly $S_0+S_1$, as computed in the previous sections. 
The interplay between these two contributions has a diagrammatic representation.
In Fig.~\ref{Phi4} (left) is the $\mathcal{O}(\lambda)$ contribution to the vacuum entanglement entropy, while Fig.~\ref{Phi4} (right) is the $\mathcal{O}(\lambda)$ renormalization of the mass.

At higher loops the mass renormalization is more involved since the renormalized mass acquires dependence on the renormalization scale. However, as we are studying the ground state, then presumably only the $p_r\to0$ limit of the renormalized mass is important.
This suggests that the leading contribution to the entropy in the full theory is to take the free theory result with the bare mass replaced by the renormalized mass, using $p_r=0$ as the renormalization scale. To test this one could either go to $O(\lambda^2)$ in the loop expansion in $\lambda\phi^4$ theory, or consider an alternate theory with momentum dependence arising at the leading loop order, such as $g\phi^3$ theory, as we now explore.

\section{Cubic Interaction}\label{Cubic}

\begin{figure}[t]
\center{\includegraphics[width=3cm]{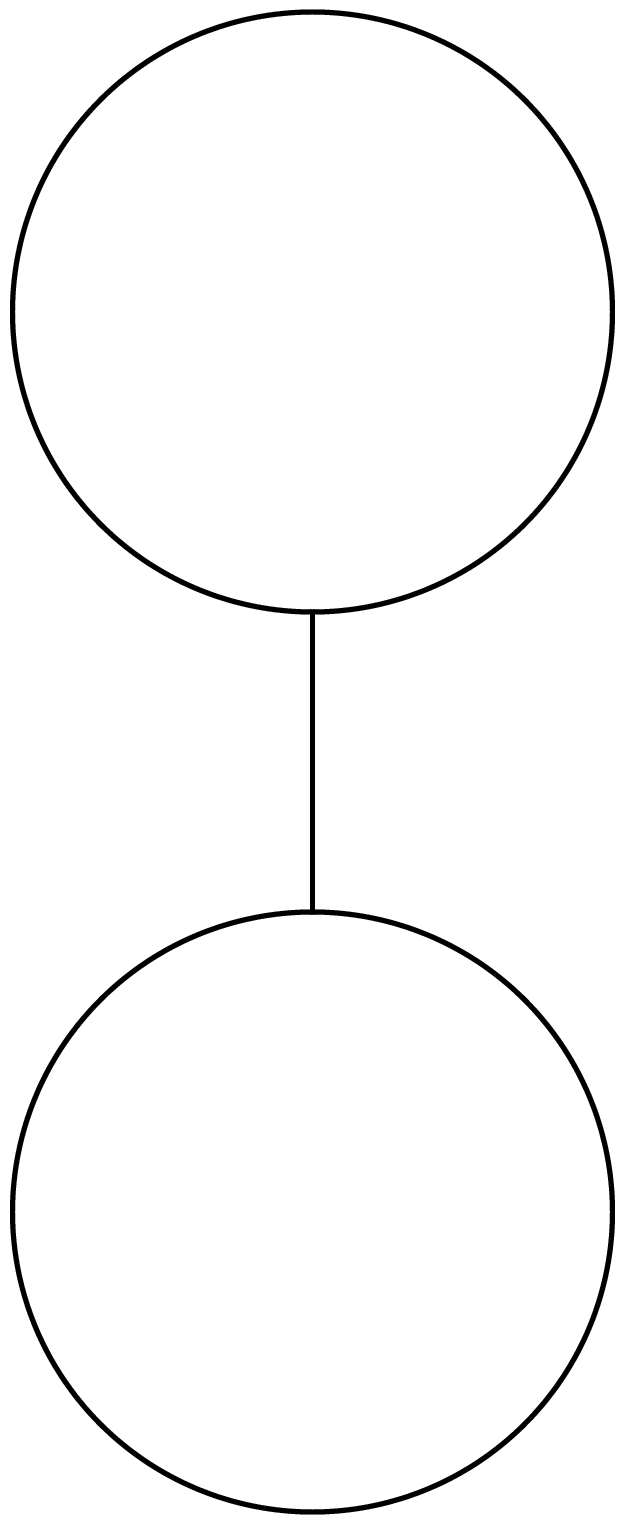}\,\,\,\,\,\,\,\,\,\,\,\,\,\,\,\,\,\,\,\,\,\,\,\,\,\,\,\,\,\,\,\,\,\,\,\,\,\,\,\,\,\,\,\,\,\,
\includegraphics[width=3cm]{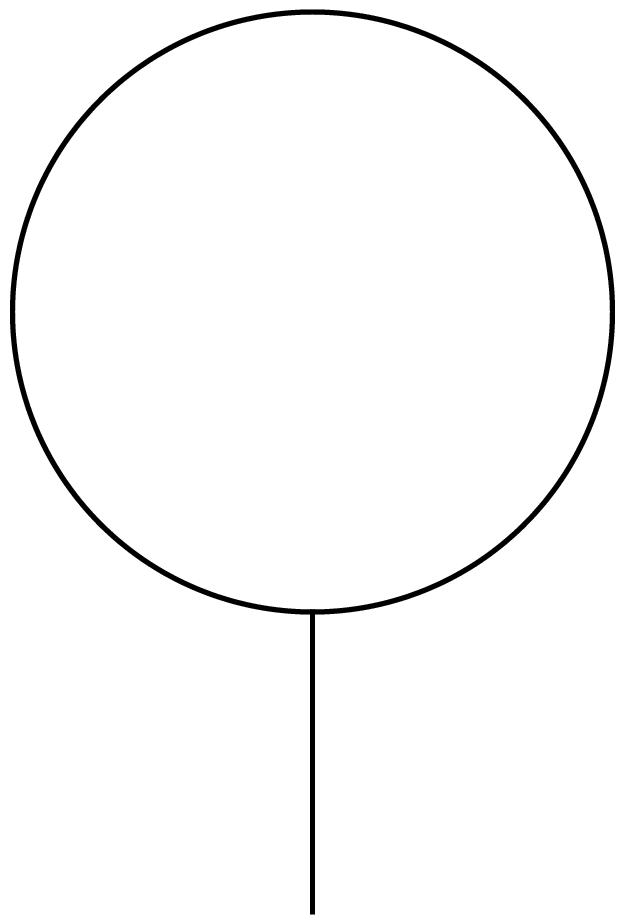}}
\caption{Tadpole diagrams for $\phi^3$ theory at order $g^2$: two-loop vacuum contribution to entropy (left) and one-loop contribution to the one point function (right).}
\label{Tadpole}\end{figure}

We now consider a field theory with a cubic interaction of the form
\beq
\mathcal{L}={1\over2}(\partial\phi)^2-{1\over2}m^2\phi^2-{g\over3!}\phi^3
\eeq
In this case the partition function can be expanded to $\mathcal{O}(g^2)$ as
\bea
\ln\mathcal{Z}_n \amp=\amp \ln\mathcal{Z}_{n,0}+{g^2\over2!(3!)^2}\int_n d^D\!x\!\int_n d^D\!x'
\left\langle \phi({\bf x})^3\phi({\bf x}')^3\right\rangle_0+\ldots\\
\amp=\amp \ln\mathcal{Z}_{n,0}+{g^2\over2!(3!)^2}\int_n d^D\!x\!\int_n d^D\!x'
\Big[6\,G_n({\bf x},{\bf x}')^3\nonumber\\
\amp\amp \,\,\,\,\,\,\,\,\,\,\,\,\,\,\,\,\,\,\,\,\,\,\,\,\,\,\,\,\,\,\,\,\,\,\,\,\,\,\,\,\,\,\,\,
+9\,G_n({\bf x},{\bf x})G_n({\bf x},{\bf x}')G_n({\bf x}',{\bf x}')\Big]+\ldots
\label{Zphi3}\eea
The second term in the integral represents a tadpole contribution. As is well known, in order to consistently expand around $\phi=0$ in the quantum theory, we must introduce a linear term $\kappa\phi$ to the Lagrangian (though it would be absent classically).
At one-loop the bare parameter $\kappa$ is required to take the value
\beq
\kappa=-{g\over 2}G_1(0)
\eeq
Once this is enforced, then all tadpole contributions drop out. This can be understood as a cancellation between 
the tadpole contribution to the vacuum entanglement entropy depicted in Fig.~\ref{Tadpole} (left) and
the generation of a one-point function depicted in Fig.~\ref{Tadpole} (right).

\begin{figure}[t]
\center{\includegraphics[width=3.3cm]{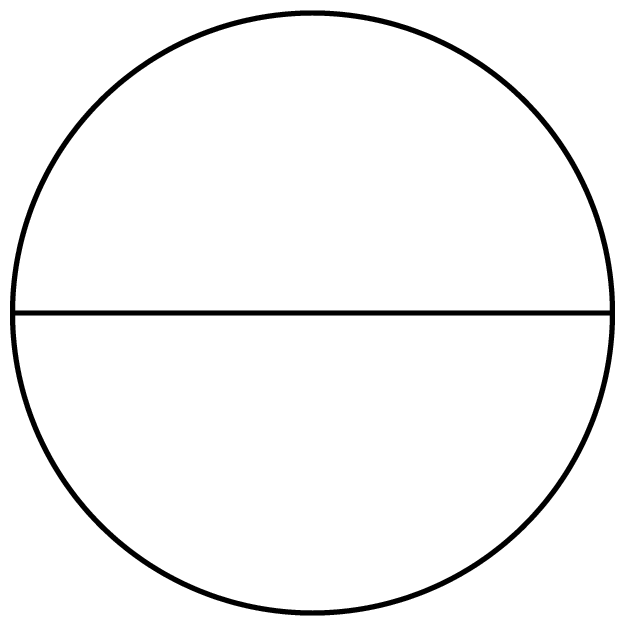}\,\,\,\,\,\,\,\,\,\,\,\,\,\,\,\,\,\,\,\,\,\,\,\,\,\,\,\,\,\,\,\,\,\,\,\,\,\,\,
\includegraphics[width=6.6cm]{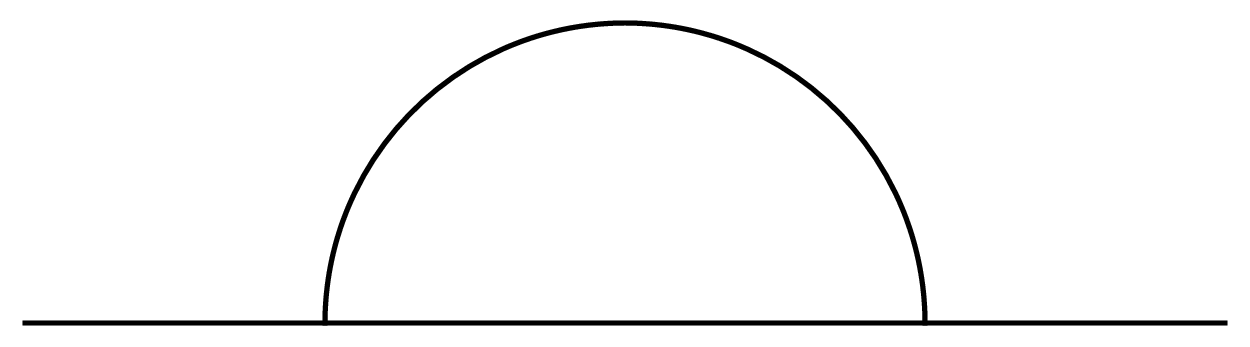}}
\caption{Feynman diagrams for $\phi^3$ theory at order $g^2$: two-loop vacuum contribution to entropy (left) and one-loop contribution to mass renormalization (right).}
\label{Phi3}\end{figure}

On the other hand, the first term in the integral eq.~(\ref{Zphi3}) is very important and is connected to the mass renormalization. The corresponding contribution to the density matrix is
\bea
\ln\mbox{Tr}[\rho^n_{\,1}]={6\,g^2\over2!(3!)^2}\left[\int_n d^D\!x\!\int_n d^D\!x'\, G_n({\bf x},{\bf x}')^3
-n\!\int d^Dx\!\int d^D\!x'\, G_1(|{\bf x}-{\bf x}'|)^3\right]
\label{Phi3Den}\eea
whose diagrammatic form is given in Fig.~\ref{Phi3} (left).

By construction the Green's function can be formally decomposed into a flat space contribution that carries translational invariance and a correction that does not. We write this as
\beq
G_n({\bf x},{\bf x}')=G_1(|{\bf x}-{\bf x}'|)+f_n({\bf x},{\bf x}')
\label{GreenNoCoinc}
\eeq
which is a natural generalization of the decomposition we performed earlier in eq.~(\ref{GreenCoinc}).  
Here we will not need to write out the full details of $f_n$ (it can be inferred from the Euler MacLaurin formula), but we will specify some specific properties.
In particular, away from the tip of the cone, $f_n$ is finite as ${\bf x}'\to{\bf x}$ since the divergence is captured by the flat space divergence $G_1$.

We insert (\ref{GreenNoCoinc}) into (\ref{Phi3Den}) and expand out the cubic piece as $G_n^3=G_1^3+3G_1^2f_n+3G_1f_n^2+f_n^3$. 
The first term in this expansion $\sim G_1^3$ is naturally paired with the second term in (\ref{Phi3Den}), leading to the following difference
\bea
{6\,g^2\over2!(3!)^2}\left[\int_n d^D\!x\!\int_n d^D\!x'\, G_1(|{\bf x}-{\bf x}'|)^3
-n\!\int\! d^Dx\!\int\! d^D\!x'\, G_1(|{\bf x}-{\bf x}'|)^3\right]
\label{Phi3DenL}\eea
By defining ${\bf y}={\bf x}-{\bf x}'$ and using translational invariance, the second term is
\beq
\sim n\, V \int\! d^D\!y\,G_1(y)^3
\eeq
where $V$ is the (flat) space-time volume. On the other hand, the first term is different since it involves a double integral over the cone.
However, the divergent part, which arises from the region ${\bf x}'\to{\bf x}$, is proportional to the volume of the cone $\sim n\,V$. This leads to a natural cancellation between these two volume divergences.

Instead the leading contribution to the entropy at $\mathcal{O}(g^2)$ is from the $3G_1^2f_n$ piece, leading to
\beq
\ln\mbox{Tr}[\rho^n_{\,1}]={18\,g^2\over2!(3!)^2}\int_n d^D\!x\!\int_n d^D\!x'\, G_1(|{\bf x}-{\bf x}'|)^2 f_n({\bf x},{\bf x}')
\eeq
We can readily extract the leading contribution to this from studying the ${\bf x}'\to{\bf x}$ regime. This gives the result
\bea
\ln\mbox{Tr}[\rho^n_{\,1}]\amp=\amp{18\,g^2\over2!(3!)^2}\int d^D\!y\, G_1(y)^2\!\int_n d^D\!x\, f_n(r)\\
\amp=\amp{18\,g^2\over2!(3!)^2}{1-n^2\over 12n}\left(\int \! d^D\!y\, G_1(y)^2\right)\!\int{d^{d_\perp}p_\perp\over(2\pi)^{d_\perp}}{1\over m^2+p_{\perp}^2}
\eea
where we have used $f_n(r)=f_n({\bf x},{\bf x})$, as given earlier in eqs.~(\ref{GreenCoinc},\,\ref{GreenCoinc2}).
Note that the $n$ subscript on the integral over $y$ has been dropped, since the divergence structure is away from the tip of the cone (except for a set of measure zero).

Then by evaluating the derivative with respect to $n$ at $n=1$ and combining with $S_0$, we obtain the following result for the entanglement entropy
\bea
S\amp=\amp S_0+S_1+\ldots\\
\amp=\amp-{1\over 12}A_\perp \int\!{d^{d_\perp}\!p_\perp\over(2\pi)^{d_\perp}}
\left[\ln(m^2+p_\perp^2)+{g^2\over 2}\left(\int \! d^D\!y\, G_1(y)^2\right){1\over m^2+p_\perp^2} \right]+\mbox{const}+\ldots\,\,\,\,\label{ScubRes}
\eea

For $d=3$ the factor $\int d^Dy\,G_1(y)^2$ gives a logarithmic divergence, leading to a contribution to the entropy of the form $S_1\sim g^2A\ln\epsilon$ (multiplied by a secondary logarithmic factor), as anticipated in the introduction.
However, as in the case of $\lambda\phi^4$ theory, this result can be recast in the form of mass renormalization.
From Fig.~\ref{Phi3} (right) the one-loop mass renormalization in $g\phi^3$ theory at renormalization scale of zero momentum is
\bea
\delta m^2\amp=\amp{g^2\over 2}\!\int\! {d^D p\over(2\pi)^D}\hat{G}_1(p)\hat{G}_1(-p)\\
\amp=\amp {g^2\over 2}\!\int\! d^D\!y\, G_1(y)^2
\eea
For $d\leq 3$ we can approximate $m_r^2=m^2+\delta m^2$, since wave-function renormalization only induces finite corrections at this order 
(see the next section for a related comment on this in the case of counterterms).
By substituting this into eq.~(\ref{FullEntropy}) and then Taylor expanding to $\mathcal{O}(g^2)$, we recover the result for the entropy as given in eq.~(\ref{ScubRes}).
So we have seen a consistent canceling of divergences, leading to eq.~(\ref{FullEntropy}). This gives the leading order result for the entropy as simply the free theory result with the bare mass replaced by the renormalized mass evaluated at $p_r=0$.


\section{Discussion}\label{Discussion}

We have shown that the UV cutoff independent area law that exists in the free theory persists in the interacting theory through a renormalization of the mass.
This means that as we vary the IR parameters, the quadratically divergent contribution $S_{div}\sim A/\epsilon^2$ (for $d=3$) cancels, leaving a UV cutoff independent area law, given in eq.~(\ref{FullEntropy}). This was determined by using the replica trick, requiring analysis on a conical geometry. Renormalization on the cone is non-trivial, due to the curvature at the tip of the cone, however the space-time volume divergences cancel in the entropy (this is true for both the entanglement entropy, as well as the Renyi or Tsallis entropies, etc) due to an interplay between two-loop vacuum contributions to the entropy and one-loop flat space mass renormalization.
(It is common for a vacuum diagram at loop order $L$ to be related to certain $N$-point functions at loop order $L-1$; for another scalar field example of this see \cite{Hertzberg:2010yz}).

The renormalization is readily seen in terms of the bare fields/couplings, as we have demonstrated here. However, we have also performed the calculation and obtained the same results with the Lagrangian re-written in terms of renormalized fields/couplings and counterterms. In the latter case, a new space-time volume divergence emerges in the quadratic sector due to wave-function renormalization, but is found to cancel against a corresponding counterterm.

Much recent discussion of entanglement entropy has focussed on conformal field theories. These works are particularly interesting at a critical point, say a quantum phase transition, since systems generally organize themselves into conformal field theories at such a critical point. Presumably, finite contributions to the entropy at a quantum phase transition encodes some information about the universality class that the field theory lies in.
In such cases the entropy has a divergent piece $S_{div}\sim A/\epsilon^2$ (for $d=3$) and the finite pieces are $\mathcal{O}(1)$ since there is no scale present, by definition, in order to have a finite area law. On the other hand, by introducing a mass, as we have done in this work, we obtain $\Delta S\sim A\, m_r^2 \ln m_r^2$. Although this explicitly breaks the scale invariance, it is of interest {\em near} the critical point. The renormalized mass $m_r$ varies in a singular fashion near the critical point (singular in the sense that it's derivatives diverge), and hence we are extracting a singular contribution to the entropy. We have in mind the following hierarchy: $\epsilon^2\ll 1/m_r^2\ll A$, i.e., the correlation length $\xi\sim 1/m_r$ is much larger than the inter-particle spacing $\epsilon$, as would be the case {\em near} a critical point. But, at the same time, the characteristic macroscopic size of the system $\sqrt{\!A}$ can still be much larger than this correlation length. In this regime we have $A\,m_r^2\gg 1$, so the UV cutoff independent area law contribution to the entropy can be numerically large, and conceivably measurable \cite{Klich:2008un}.

Simple models of symmetry breaking involve a scalar field interacting under a potential $V(\phi) = -{\mu^2\over2}\sigma^2+{\lambda\over4!}\sigma^4$. Expanding around the vacuum (whether the symmetry is broken $\mu^2>0$ or unbroken $\mu^2<0$) leads to a potential that fits into the form analyzed in this paper eq.~(\ref{Lagrangian}), for specific values of $g,m$. Hence, one can apply the cutoff independent entropy law to symmetry breaking models, including application to Ginzburg-Landau type treatments of phase transitions, which would be of interest.

It is important to extend these results, which may be possible in several directions. Firstly, here we have focussed on a half space geometry, though other geometries are of interest. For instance, one may consider a waveguide, as we studied earlier in Ref.~\cite{Hertzberg:2010uv}, and perform some expansion in powers of area, perimeter, curvature, etc (perhaps similar to \cite{Hertzberg:2005pr,Hertzberg:2007ch}), or one may consider finite intervals or spheres, etc. Secondly, another important extension is to consider other interactions and other fields, fermions or gauge bosons (e.g.,  \cite{Casini:2005a,Buividovich:2008gq,Velytsky:2008rs}).
Finally, it is of interest to extend the results to higher loop order in the perturbative expansion, as well as tracking other finite sub-leading corrections.

\section*{Acknowledgments}
I would like to thank Shamik Banerjee, Erik Tonni, and Frank Wilczek for very helpful discussions.
I would like to acknowledge support from the Center for Theoretical Physics at MIT where the later part of this work was performed,
and SITP and KIPAC at Stanford where the early part of this work was performed.
This work is supported by the U.S. Department of Energy under cooperative research agreement Contract Number DE-FG02-05ER41360,
as well as NSF grant PHY-0756174.

\end{document}